\documentclass[prd,aps,preprintnumbers, showpacs, nofootinbib,superscriptaddress,notitlepage]{revtex4-1}

\usepackage{epsfig}
\usepackage{bm}
\usepackage{amssymb}
\usepackage{amsmath}
\usepackage{color}
\usepackage{subfigure}
\usepackage[pdfstartview=FitH]{hyperref}
\hypersetup{colorlinks=true, citecolor=blue, linkcolor=blue,filecolor=black,urlcolor=blue}

\begin{document}

\title{On the Dihadron Angular Correlations in Forward $pA$ collisions}

\author{Anna Stasto}

\affiliation{Department of Physics, The Pennsylvania State University, University Park, PA 16802, United States}

\author{Shu-Yi Wei}
\affiliation{Key Laboratory of Quark and Lepton Physics (MOE) and Institute
of Particle Physics, Central China Normal University, Wuhan 430079, China}
\affiliation{Centre de Physique Th\'eorique, Ecole Polytechnique, 
CNRS, Universit\'e Paris-Saclay, Route de Saclay, 91128 Palaiseau, France.}

\author{Bo-Wen Xiao}
\affiliation{Key Laboratory of Quark and Lepton Physics (MOE) and Institute
of Particle Physics, Central China Normal University, Wuhan 430079, China}
\affiliation{Centre de Physique Th\'eorique, Ecole Polytechnique, 
CNRS, Universit\'e Paris-Saclay, Route de Saclay, 91128 Palaiseau, France.}

\author{Feng Yuan}
\affiliation{Nuclear Science Division, Lawrence Berkeley National
Laboratory, Berkeley, CA 94720, USA}


\begin{abstract}
Dihadron angular correlations in forward 
$pA$ collisions have been considered as one of the most sensitive observables to the gluon saturation effects. 
In general, both parton shower effects and saturation effects are responsible for the back-to-back dihadron angular de-correlations. With the recent progress in the saturation formalism, 
we can incorporate the parton shower effect by adding the corresponding Sudakov factor in the saturation framework. In this paper, we carry out the first detailed numerical study in this regard, and 
find a very good agreement with previous RHIC $pp$ and $dAu$ data. 
This study can help us to establish a baseline in $pp$ collisions which contains little saturation effects, and further make predictions for dihadron angular correlations in $pAu$ collisions, which will allow to  search for the signal of parton saturation. 
\end{abstract}
\maketitle


\section{Introduction} 

Small-$x$ physics framework provides with the description of dense parton densities at high energy limit, when the longitudinal momentum fraction $x$ of partons with respect to parent hadron is small. It predicts the onset of the gluon saturation phenomenon\cite{Gribov:1984tu, Mueller:1985wy, Gelis:2010nm} as a result of nonlinear QCD evolution\cite{BK, JIMWLK} when the gluon density becomes very high.

Dihadron angular decorrelation in forward rapidity $pA$ collisions, which was first proposed in Ref.~\cite{Marquet:2007vb}, is reckoned as one 
of the most interesting observables sensitive to gluon saturation effects. There have been great
 theoretical~\cite{Marquet:2007vb, Albacete:2010pg, Dominguez:2010xd, Dominguez:2011wm, Stasto:2011ru, Lappi:2012nh, Iancu:2013dta} and experimental~\cite{Braidot:2010ig, Li:2011we, Adare:2011sc} efforts devoted to this topic over the last few years. In addition, by applying the small-$x$ improved TMD factorization framework~\cite{Kotko:2015ura}, the suppression of the forward dijet angular correlations in proton-lead versus proton-proton collisions at the LHC due to saturation effects has been predicted in Ref.~\cite{vanHameren:2014lna, vanHameren:2016ftb}. 
Besides the calculations based on the saturation formalism, there are also other explanations based on the cold nuclear matter energy loss effects and coherent power corrections, as shown in Refs.~\cite{Qiu:2004da, Kang:2011bp}.

More precise data on the dihadron angular correlations in the forward rapidity region in $pAu$ collisions from the STAR collaboration at RHIC are expected to be released soon. The prediction due to the saturation effect shows clear enhancement of decorrelations in $pAu$
collisions as compared to that in $pp$ collisions. The new data will also allow  to examine the strength of saturation effects in different $p_T$ bins and conduct detailed comparison between the experimental data and theoretical predictions. In addition, the pedestal due to double parton distributions observed in $dAu$ collisions, which is considered to be a background, is expected to be much smaller in forward $pAu$ collisions. 

On the theory side, recent developments have allowed  to incorporate the so-called parton shower effect, namely the Sudakov effect, into the small-$x$ 
formalism~\cite{Mueller:2013wwa, Sun:2014gfa, Mueller:2016xoc}. This, in particular, will enable us to go beyond the saturation dominant region, 
and conduct calculations for dihadron correlation in a much wider regime where both saturation effects and Sudakov effects 
are important. Thus, we can perform a much more comprehensive and quantitative comparison between the small-$x$ calculation and experimental data. In general, both saturation  and Sudakov effect should play important roles in dihadron (dijet) angular correlation (decorrelation) in $pAu$ collisions. Furthermore, similar technique has been applied to dijet and dihadron productions in the central rapidity region in both $pp$ and heavy ion collisions~\cite{Mueller:2016gko, Chen:2016vem}. It has been demonstrated to be useful in the study of the transport coefficient of the quark-gluon plasma by comparing the angular correlations in $pp$ and $AA$ collisions. 
The Sudakov effects have also been incorporated in the recent calculation of the forward dijet production in ultraperipheral heavy ion collisions at LHC~\cite{Kotko:2017oxg}. The calculation was based on the framework that interpolates between the Color Glass Condensate formalism and high energy factorization. The Sudakov effects have been included by the suitable re-weighting procedure of the events  using the Sudakov form factor in a Monte Carlo simulation.

In Ref.~\cite{Mueller:2013wwa, Mueller:2016xoc}, it has been demonstrated that the small-$x$ effects and Sudakov effects can be simultaneously taken into account in the auxiliary $b_\perp$ space as a result of convolutions in the momentum space. Saturation effect in forward $pAu$ collisions can be factorized into various small-$x$ unintegrated gluon distributions (UGDs) as derived in Refs.~\cite{Dominguez:2010xd, Dominguez:2011wm}.  
These UGDs include two important ingredients of saturation physics, namely small-$x$ (non-linear) evolution and multiple interaction, which can be characterized by the saturation momentum $Q_s^2 (x_g)$ and products of several scattering amplitudes (including both quadrupole and dipole type). Generally speaking, one expects that the saturation effect is stronger in the region where the gluon momentum fraction $x_g$ becomes smaller. This implies that the saturation effect is maximized in the lowest $p_T$ bin of dihadrons at given rapidity. On the other hand, the strength of the Sudakov effect depends on the hardness of the scattering, namely the magnitude of $p_T$ of each jet prior to the fragmentation process. Therefore, one expects that the parton shower effect is relatively weaker in the low $p_T$ bins while it grows stronger for large $p_T$ bins. In dijet productions, we have learnt that the angular correlation of dijets in $pp$ collisions always becomes steeper for dijets with larger jet transverse momenta. 
Therefore, we expect that dihadrons in high $p_T$ bins are more sharply correlated (steeper) than those low $p_T$ bins, since the saturation effects become weaker in high $p_T$ bins while Sudakov effects only grows slowly with increased $p_T$. As a result, we can expect that the curves of back-to-back dihadron angular correlation become more and more flat when one moves from large $p_T$ bins to small $p_T$ bins. The purpose of this paper is to conduct a comprehensive phenomenological study on the dihadron angular correlations by comparing with all the available data and making predictions for upcoming data. 

To take into account the small-$x$ effect, we use the simple Golec-Biernat Wusthoff (GBW) model~\cite{GolecBiernat:1998js} as a first step, since it is easy to implement and at the same time contains relevant physics due to the saturation. In principle, one should use a more sophisticated approach which employs the solution~\cite{Marquet:2016cgx} to the non-linear small-$x$ evolution equations~\cite{BK, JIMWLK, Dominguez:2011gc, Dumitru:2011vk} for various types of gluon distributions to compute the correlation as in Ref.~\cite{futurestudy}. The is much more numerically demanding together with the Sudakov resummation. Therefore, we will leave this for a future work.

This paper is organized as follows. In Sec. II, we provide the summary of the theoretical formulas for dihadron productions in the forward rapidity region and discuss details of  the numerical implementation of the Sudakov factor in the small-$x$ formalism. In Sec. III, we show the comparison between our numerical result with the experimental data measured at RHIC and provide our prediction for the upcoming data in $pAu$ collisions. We summarize our findings in Sec. IV.


\section{Forward Rapidity Dihadron Production in pA collisions} 

Following Ref.~\cite{Dominguez:2010xd, Dominguez:2011wm, Stasto:2011ru}, we study the forward dihadron production in the so-called hybrid dilute-dense factorization, which is motivated by the fact that the projectile proton is dilute while the target nucleus (or proton) is rather dense in such kinematical region. For the quark initiated channel, the back-to-back dihadron production formula can be written as the convolution of the large $x$ collinear quark distribution from the projectile proton, the small-$x$ UGDs from the target nucleus, and the hard factor as well as the final state fragmentation functions as follows
\begin{eqnarray}
 \frac{d\sigma^{qg\to gq\to h_1 h_2}}{dy_1 dy_2 d^2 p_{1\perp}d^2 p_{2\perp}} &=& \int \frac{dz_1}{z_1^2} \int \frac{dz_2}{z_2^2} \left\{D_{h/g}(z_1)D_{h/q}(z_2) x q(x) H_{qg}\left [ (1-z)^2 \mathcal{F}_{qg}^{(a)}(x_g, q_\perp)+\mathcal{F}_{qg}^{(b)}(x_g, q_\perp)\right]\right.\nonumber\\
 && \quad \quad \quad \quad \quad \quad \left.+[1\leftrightarrow 2] \vphantom{ \mathcal{F}_{qg}^{(a)}} \right\} , \label{improved}
 \end{eqnarray}
where $P_\perp \equiv (1-z)k_{1\perp }-z k_{2\perp }$ and $q_\perp \equiv k_{1\perp }+ k_{2\perp }$ with $z=\frac{|k_{1\perp }|e^{y_1}}{|k_{1\perp }|e^{y_1}+|k_{2\perp }|e^{y_2}}$, $k_{1\perp }=p_{1\perp }/z_1$ and $k_{2\perp }=p_{\perp 1}/z_2$. We use $y_1, \, p_{\perp 1}$ and $y_2, \, p_{\perp 2}$ to represent the rapidity and transverse momenta of the trigger hadron and associate hadron, respectively. The $q(x)$ is the collinear quark distribution function. We use CT14\cite{Dulat:2015mca} from the CTEQ group in the numerical calculation. $D_{h/q}(x)$ and $D_{h/g}(x)$ are the collinear parton fragmentation functions. In the numerical evaluation, AKK08\cite{Albino:2008fy} fragmentation functions are used. The factorization scale $\mu$ is set to be $\mu_b$ (defined below) in the Sudakov resummation framework, in order to reach a convenient and compact resummation fornula. As common practice, the $b_\perp$ dependence in the factorization scale $\mu$ should also be taken into account when the numerical integration over $b_\perp$ is carried out. The hard factor $H_{qg}$ and small-$x$ gluon distributions are defined as
\begin{eqnarray}
H_{qg}&=&\frac{\alpha_s^2}{2P_\perp^4} \left[1+(1-z)^2\right] (1-z),  \\
\mathcal{F}_{qg}^{(a)}(x_g, q_\perp)&=&\frac{-N_c S_\perp}{2\pi^2 \alpha_s} \int_0^{\infty} \frac{b_\perp db_\perp}{2\pi} J_0(q_\perp b_\perp)  e^{-S_{\textrm{Sud}}^{q+g\to q+g}(b_\perp)} \nabla_{b_\perp}^2 S_{x_g}(b_\perp),\\
\mathcal{F}_{qg}^{(b)}(x_g, q_\perp)&=&\frac{C_F S_\perp}{2\pi^2 \alpha_s} \int_0^{\infty} \frac{b_\perp db_\perp}{2\pi} J_0(q_\perp b_\perp)  e^{-S_{\textrm{Sud}}^{q+g\to q+g}(b_\perp)} \frac{\nabla_{b_\perp}^2 \ln\tilde{S}_{x_g}(b_\perp)}{\ln\tilde{S}_{x_g}(b_\perp)} \left[1-\tilde{S}_{x_g}(b_\perp)\right] S_{x_g}(b_\perp).
\end{eqnarray}
Here we denote $S_{x_g}(b_\perp)$ and $\tilde{S}_{x_g}(b_\perp)$ as the small-$x$ expectation value of fundamental and adjoint Wilson loops with space separation $b_\perp$, respectively. $S_\perp$ is denoted as the averaged transverse area of the target hadron. In principle, besides the dipole amplitude, quadrupole scattering amplitudes also appear in the production of dihadrons as demonstrated in Ref.~\cite{Dominguez:2010xd, Dominguez:2011wm}. We have used the so-called dipole approximation to write the quadrupole amplitude in terms of dipole amplitudes in the adjoint representation. For the gluon initiated channel, the corresponding cross section is 
\begin{eqnarray}
\frac{d\sigma^{gg\to gg\to h_1 h_2}}{dy_1 dy_2 d^2 p_{\perp1}d^2 p_{\perp2}} &=& \int \frac{dz_1}{z_1^2} \int \frac{dz_2}{z_2^2} D_{h/g}(z_1)D_{h/g}(z_2) x g(x) H_{gg} \nonumber\\
&& \times  \left \{ \left[z^2+(1-z)^2\right ] \mathcal{F}_{gg}^{(a)}(x_g, q_\perp)+2z(1-z)\mathcal{F}_{gg}^{(b)}(x_g, q_\perp)+\mathcal{F}_{gg}^{(c)}(x_g, q_\perp) \right\},  \label{gg}
\end{eqnarray}
where the hard factor $H_{gg}$ and small-$x$ gluon distributions are 
\begin{eqnarray}
H_{gg}&=&\frac{2\alpha_s^2}{P_\perp^4} \left[1-z(1-z)\right]^2 ,  \\
\mathcal{F}_{gg}^{(a)}(x_g, q_\perp)&=&\frac{-N_c S_\perp}{2\pi^2 \alpha_s} \int_0^{\infty} \frac{b_\perp db_\perp}{2\pi} J_0(q_\perp b_\perp)  e^{-S_{\textrm{Sud}}^{g+g\to g+g}(b_\perp)}S_{x_g}(b_\perp) \left[ \nabla_{b_\perp}^2 S_{x_g}(b_\perp)\right],\\
\mathcal{F}_{gg}^{(b)}(x_g, q_\perp)&=&\frac{N_c S_\perp}{2\pi^2 \alpha_s} \int_0^{\infty} \frac{b_\perp db_\perp}{2\pi} J_0(q_\perp b_\perp)  e^{-S_{\textrm{Sud}}^{g+g\to g+g}(b_\perp)} \left[\nabla_{b_\perp} S_{x_g}(b_\perp)\right] \cdot\left[\nabla_{b_\perp} S_{x_g}(b_\perp)\right],\\
\mathcal{F}_{gg}^{(c)}(x_g, q_\perp)&=&\frac{C_F S_\perp}{2\pi^2 \alpha_s} \int_0^{\infty} \frac{b_\perp db_\perp}{2\pi} J_0(q_\perp b_\perp)  e^{-S_{\textrm{Sud}}^{g+g\to g+g}(b_\perp)} \frac{\left[\nabla_{b_\perp}^2 \ln\tilde{S}_{x_g}(b_\perp)\right]}{\ln\tilde{S}_{x_g}(b_\perp)} \left[1-\tilde{S}_{x_g}(b_\perp)\right] S_{x_g}(b_\perp) S_{x_g}(b_\perp).
\end{eqnarray}
We have also computed the $gg\to q\bar q$ channel, which is found to be always negligible numerically. If the corresponding Sudakov factors $S_{\textrm{Sud}}(b_\perp)$ are set to be zero, the above expressions reduce to the results originally derived in Refs.~\cite{Dominguez:2010xd, Dominguez:2011wm} and numerically evaluated in Ref.~\cite{Stasto:2011ru}. The Sudakov factors come from the resummation of soft-collinear gluon radiation and they can be normally written as follows
\begin{equation}
S_{\textrm{Sud}}^{a+b\to c+d} (b_\perp) =\sum_{i=a,b,c,d} S_p^{i} (b_\perp) + \sum_{i=a, c, d}S^{i}_{np} (b_\perp),
\end{equation}
where $S_p^{i} (b_\perp) $ and $S_{np}^{i} (b_\perp)$ are the perturbative and non-perturbative Sudakov factors, respectively for parton $i$. Since we are using small-$x$ unintegrated gluon distributions for parton $b$, which may have already contained some non-perturbative information at low-$x$ about the target nuclei(protons), we do not include non-perturbative Sudakov factor associated with the incoming small-$x$ gluon (active parton b) in $S^{i}_{np} $. In addition,
according to the derivation in Ref.~\cite{Mueller:2013wwa}, the single logarithmic term, which is known as the $B$-term, in the perturbative part of the Sudakov factor for this incoming small-$x$ gluon is absent. The perturbative Sudakov factors for $q+g\to q+g$ and $g+g\to g+g$ channels are given by
\begin{eqnarray}
S_p^{q+g\to q+g} (Q, b_\perp) & = & \int_{\mu_b^2}^{Q^2} \frac{d\mu^2}{\mu^2} \left[
2 (C_F + C_A) \frac{\alpha_s}{2\pi} \ln \left( \frac{Q^2}{\mu^2} \right)
- \left(\frac{3}{2}C_F +  C_A \beta_0 \right) \frac{\alpha_s}{\pi} \right],\\
S_p^{g+g\to g+g} (Q, b_\perp) & = & \int_{\mu_b^2}^{Q^2} \frac{d\mu^2}{\mu^2} \left[
  4 C_A \frac{\alpha_s}{2\pi} \ln \left( \frac{Q^2}{\mu^2} \right)
- 3 C_A \beta_0 \frac{\alpha_s}{\pi} \right].
\end{eqnarray}
$\beta_0 = (11-2n_f/3)/12$, $\mu_b=2e^{-\gamma_E}/b_*$, and $b_* = b_\perp/\sqrt{1+b_\perp^2/b_{\rm max}^2}$. For the non-perturbative Sudakov factor, we employ the parameterization in \cite{Su:2014wpa,Prokudin:2015ysa}.
\begin{eqnarray}
S_{np}^{q+g\to q+g} (Q, b_\perp) & = & \left( 2 + \frac{C_A}{C_F} \right) \frac{g_1}{2} b_\perp^2 + \left(2 + \frac{C_A}{C_F} \right) \frac{g_2}{2} \ln \frac{Q}{Q_0} \ln \frac{b_\perp}{b_*}, \\
S_{np}^{g+g\to g+g} (Q, b_\perp) & = & \frac{3C_A}{C_F} \frac{g_1}{2} b_\perp^2 + \frac{3C_A}{C_F} \frac{g_2}{2} \ln \frac{Q}{Q_0} \ln \frac{b_\perp}{b_*}.
\end{eqnarray}
$g_1 = 0.212$, $g_2 = 0.84$, and $Q_0^2 = 2.4 {\rm GeV}^2$. As found in Ref.~\cite{Mueller:2013wwa, Sun:2014gfa, Mueller:2016xoc}, in order to get rid of terms associated with collinear gluon splittings, it is most convenient to set the factorization scale $\mu=\mu_b$ for both collinear parton distributions and fragmentation functions in the resummed formula. 
Since we have arbitrary number of soft gluons resummed into the Sudakov factor, it becomes difficult to recover the exact kinematics. In practice\cite{Mueller:2016gko, Chen:2016vem}, we can approximately write $x=\frac{k_\perp}{\sqrt{s}} \left(e^{y_1}+e^{y_2}\right)$ and $x_g=\frac{k_\perp}{\sqrt{s}} \left(e^{-y_1}+e^{-y_2}\right)$ with $k_\perp\equiv \textrm{max}[k_{1\perp}, k_{2\perp}]$. In addition, the hard scale is then determined as $Q^2=xx_g s= k_\perp^2 \left(2+e^{y_1-y_2}+e^{y_2-y_1}\right)$. In principle, $Q$ should be much larger than the transverse momentum imbalance of the dijet pair $q_\perp \sim \frac{1}{b_\perp}>\frac{1}{b_{\rm max}}$. In the current RHIC kinematics, this is not exactly the case ($Q\sim 4$ to $10$ \textrm{GeV}), which means that the effect of the non-perturbative Sudakov factor is not completely negligible in contrast to the high energy dijet productions at the LHC\cite{Mueller:2016gko}. We are also aware of the issue of non-universality in dijet productions\cite{Collins:2007nk, Rogers:2010dm}, which implies that the non-perturbative Sudakov factors in forward dihadron productions may differ from those used in DIS or Drell-Yan processes. We rely on numerical fit in $pp$ collisions to determine the size of the non-perturbative Sudakov factors in forward dihadron production. 

As mentioned above, we employ the GBW model\cite{GolecBiernat:1998js} with Gaussian form  for the scattering amplitudes in this paper for the sake of simplicity. The $S_{x_g} (b_\perp)$ and $\tilde S_{x_g}(b_\perp)$ is then given by
\begin{align}
& S_{x_g} (b_\perp) = \exp\left(-\frac{1}{4} Q_s^2 b_\perp^2\right), \\
& \tilde S_{x_g} (b_\perp) = \exp \left( - \frac{1}{4} Q_{sg}^2 b_\perp^2 \right),
\end{align}
while, $Q_{sg}^2 = \frac{2N_c^2}{N_c^2 - 1} Q_s^2$. The various relevant gluon distributions can be then cast into
\begin{align}
& \mathcal{F}_{qg}^{(a)}(x_g, q_\perp) 
  = 
    \frac{N_c S_\perp}{2\pi^2 \alpha_s} 
  \int_0^{\infty} \frac{b_\perp db_\perp }{2\pi} J_0 (q_\perp b_\perp)
    e^{-S_{\rm sud}^{q+g\to q+g} (b_\perp)}
    Q_s^2 \left(1-\frac{1}{4}Q_s^2 b_\perp^2 \right)
    e^{-\frac{1}{4}Q_s^2 b_\perp^2}, \\
& \mathcal{F}_{qg}^{(b)}(x_g, q_\perp) 
  = 
    \frac{C_F S_\perp}{2\pi^2 \alpha_s} 
     \int_0^{\infty} \frac{b_\perp db_\perp }{2\pi} J_0 (q_\perp b_\perp)
    e^{-S_{\rm sud}^{q+g\to q+g} (b_\perp)}
    \frac{4}{b_\perp^2}
    \left(1-e^{-\frac{1}{4}Q_{sg}^2 b_\perp^2} \right)
    e^{-\frac{1}{4}Q_s^2 b_\perp^2}, \\
& \mathcal{F}_{gg}^{(a)}(x_g, q_\perp) 
  =
    \frac{N_c S_\perp}{2\pi^2 \alpha_s}
     \int_0^{\infty} \frac{b_\perp db_\perp}{2\pi} J_0 (q_\perp b_\perp)
    e^{-S_{\rm sud}^{g+g\to g+g} (b_\perp)}
    Q_s^2 \left( 1 - \frac{1}{4} Q_s^2 b_\perp^2 \right)
    e^{-\frac{1}{2} Q_s^2 b_\perp^2},\\
& \mathcal{F}_{gg}^{(b)}(x_g, q_\perp) 
  =
    \frac{N_c S_\perp}{2\pi^2 \alpha_s}
     \int_0^{\infty} \frac{b_\perp db_\perp}{2\pi} J_0 (q_\perp b_\perp)
    e^{-S_{\rm sud}^{g+g\to g+g} (b_\perp)}
    \frac{1}{4} Q_s^4 b_\perp^2
    e^{-\frac{1}{2} Q_s^2 b_\perp^2},\\
& \mathcal{F}_{gg}^{(c)}(x_g, q_\perp) 
  =
    \frac{C_F S_\perp}{2\pi^2 \alpha_s}
     \int_0^{\infty} \frac{b_\perp db_\perp}{2\pi} J_0 (q_\perp b_\perp)
    e^{-S_{\rm sud}^{g+g\to g+g} (b_\perp)}
    \frac{4}{b_\perp^2} \left(1-e^{-\frac{1}{4} Q_{sg}^2 b_\perp^2} \right)
    e^{-\frac{1}{2} Q_s^2 b_\perp^2}.
\end{align}
As shown above, the dihadron production process in the dilute-dense factorization involves several different types of gluon distribution. These distributions are related to the gluon distributions defined in inclusive DIS, however, they are in fact different type of distributions with various forms of gauge links. 





\section{Numerical results}

\begin{figure}[h!]
\includegraphics[width=0.3\linewidth]{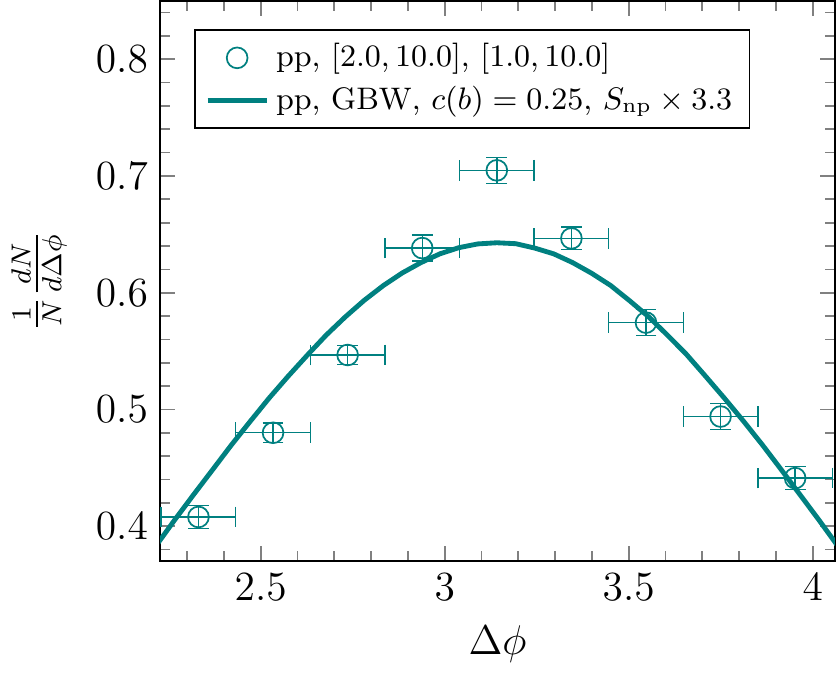}
\includegraphics[width=0.3\linewidth]{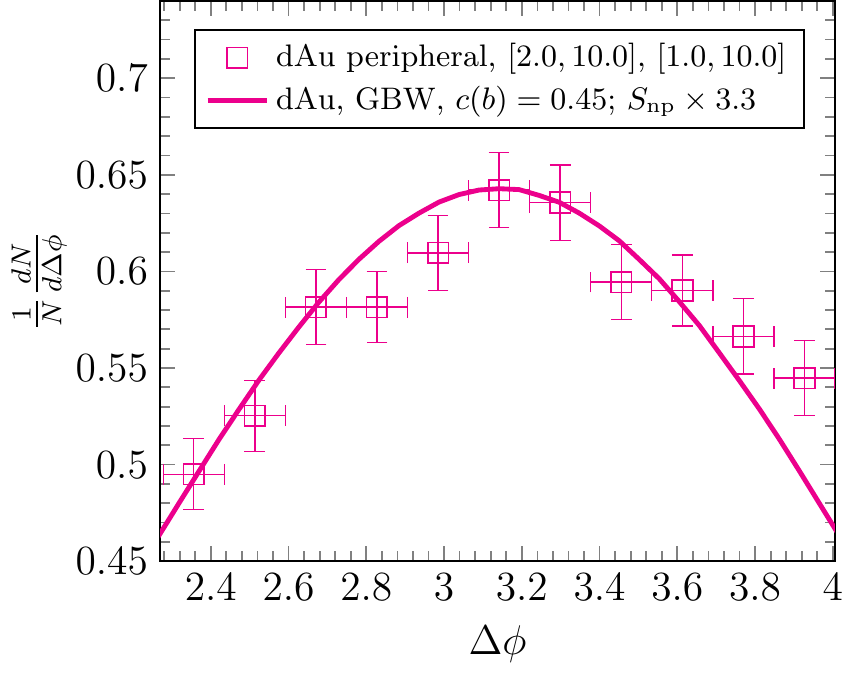}
\includegraphics[width=0.3\linewidth]{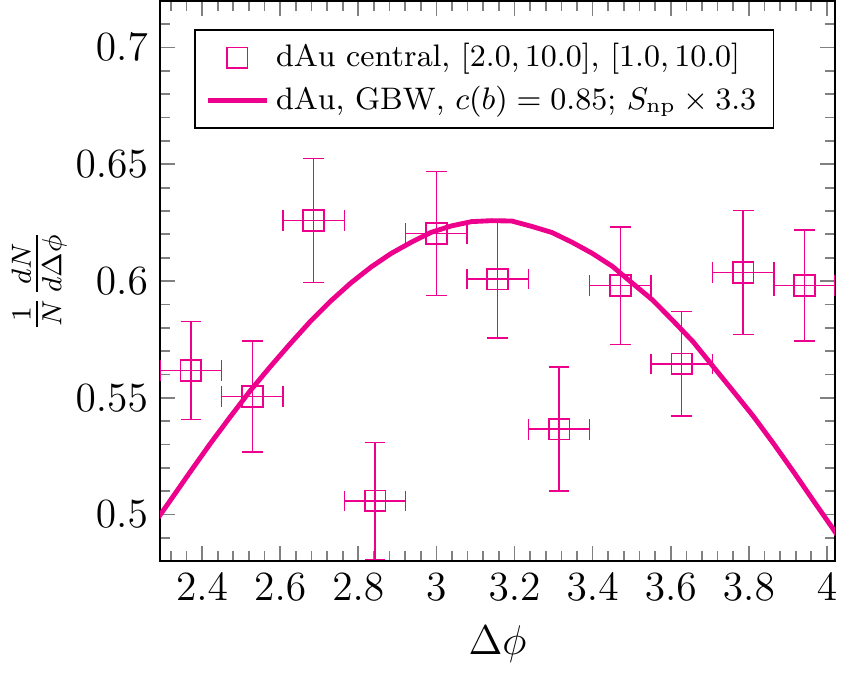}
\caption{
Normalized forward dihadron angular correlation compared  with the experimental data measure by STAR collaboration \cite{Braidot:2010ig}.
Both the leading and associate hadrons are in the forward rapidity region ($2.5<y<4$).
The pedestal has not been taken into account in the theoretical curves for the $dAu$ collisions.
}
\label{fig:forward.star}
\end{figure}

Previous experimental measurements\cite{Braidot:2010ig, Li:2011we, Adare:2011sc} and theoretical calculations\cite{Albacete:2010pg, Stasto:2011ru, Lappi:2012nh} studied the coincidence probability $C(\Delta \phi)$, which is defined as the ratio of the dihadron yield to the single trigger hadron yield. The trigger hadron yield (cross section) is used as the normalization. In this paper, we suggest to study the self-normalized angular correlation in the back-to-back region. The advantage of self-normalized correlation is that one can avoid the uncertainties and subtleties introduced by the single trigger hadron yield in the small-$x$ formalism (see for example the discussion in Ref.~\cite{Stasto:2013cha,Watanabe:2015tja, Stasto:2016wrf}). As a matter of fact, this has become the common practice at the LHC for back-to-back dijet and photon-jet angular correlation measurements. Therefore, in the following, we adopt such idea and normalize the angular correlation in the back-to-back region for both theoretical curves and experimental data. 

With the Sudakov factor, now we can not only describe the $dAu$ data in which the saturation effects are dominant, but also naturally explain width of the back-to-back correlation data measured in $pp$ collisions with $Q_{sp}^2= c(b) Q_{s, \textrm{GBW}}^2(x)$ and $c(b)=0.25$. Here we use the profile parameter $c(b)$ to take into account the fact that collisions are mostly peripheral in $pp$ collisions. Similar parametrization has been also used in single forward hadron productions in $pp$ collisions\cite{Albacete:2010bs}. The GBW saturation momentum is defined as $Q_{s, \textrm{GBW}}^2(x)\equiv \left(x/x_0\right)^{-\lambda}\textrm{GeV}^2$ with $x=3.04 \times 10^{-4}$ and $\lambda =0.288$. In addition, as explained earlier, due to the non-universality of dijet productions, we expect that the strength of the non-perturbative Sudakov factor could be different for this process. As shown in Fig.~\ref{fig:forward.star}, we find that we can explain the forward dihadron back-to-back angular correlations in $pp$ collisions with $3.3$ times of the non-perturbative Sudakov factor fitted from deep inelastic scattering (DIS) and Drell-Yan process.

Using the same parametrizations, we further perform the numerical calculation for the dihadron angular correlation in the forward rapidity region in peripheral and central $dAu$ collisions, and compare with the experimental data measured by the STAR collaboration 
\cite{Braidot:2010ig} in Fig.~\ref{fig:forward.star}.
The saturation scale in the $pA$ or $dA$ collisions is given by $Q_{sA}^2 = c(b) A^{1/3}Q_{s, \textrm{GBW}}^2$ \cite{Gelis:2010nm, GolecBiernat:1998js},
while $c(b)=0.85$ and 0.45 for the central and peripheral collisions respectively \cite{Stasto:2011ru}. For minimum bias events, we use $c(b)=0.56$ which is roughly in between the peripheral and central collision events. 

In $pp$ collisions, we find that the Sudakov and saturation effects are equally important. Therefore, the addition of the Sudakov factor is essential   to describe the back-to-back angular correlation in forward dihadron productions in $pp$ collisions in the dilute-dense factorization. In $dAu (pA)$ collisions (especially the central collisions), the saturation effects become the dominant mechanism for the broadening of the away side peak, since the saturation scale is enlarged by a factor of $A^{1/3}$ for large nuclei. Nevertheless, in order to make more reliable predictions for various transverse momentum ranges of dihadron productions, it is necessary to take into account the Sudakov effect.  

\begin{figure}[h!]
\includegraphics[width=0.35\linewidth]{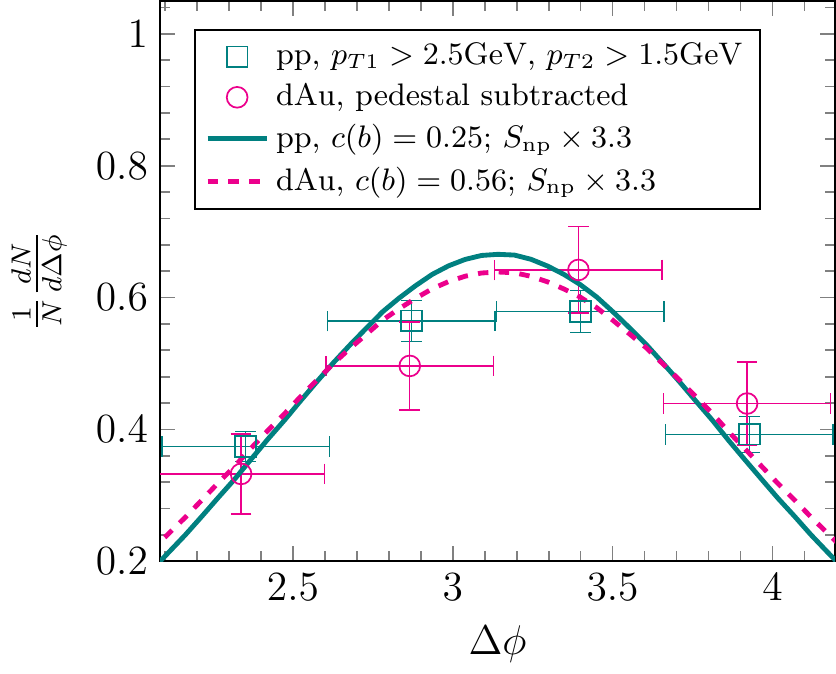}
\caption{Normalized forward and near-forward dihadron angular correlation comparing with the experimental data measure by STAR collaboration \cite{Li:2011we}.
The trigger $\pi^0$ is in the forward rapidity region ($2.5<y<4$) and the  associate $\pi^0$ is in the near forward rapidity region ($1.1<y<1.9$).
}
\label{fig:f.nf}
\end{figure}

We also perform the numerical calculation for the dihadron angular correlation in the forward and near-forward rapidity region
and compare with the experimental data \cite{Li:2011we} in Fig.~\ref{fig:f.nf}. As expected, the Sudakov effect is the dominant effect, while the small-$x$ effect is negligible 
since $x_g$ is not sufficiently small in this kinematical region. We have also checked the dihadron correlation between forward trigger hadron and middle rapidity associate hadron \cite{Li:2011we}, and find the same conclusion.

\begin{figure}[h!]
\includegraphics[width=0.35\linewidth]{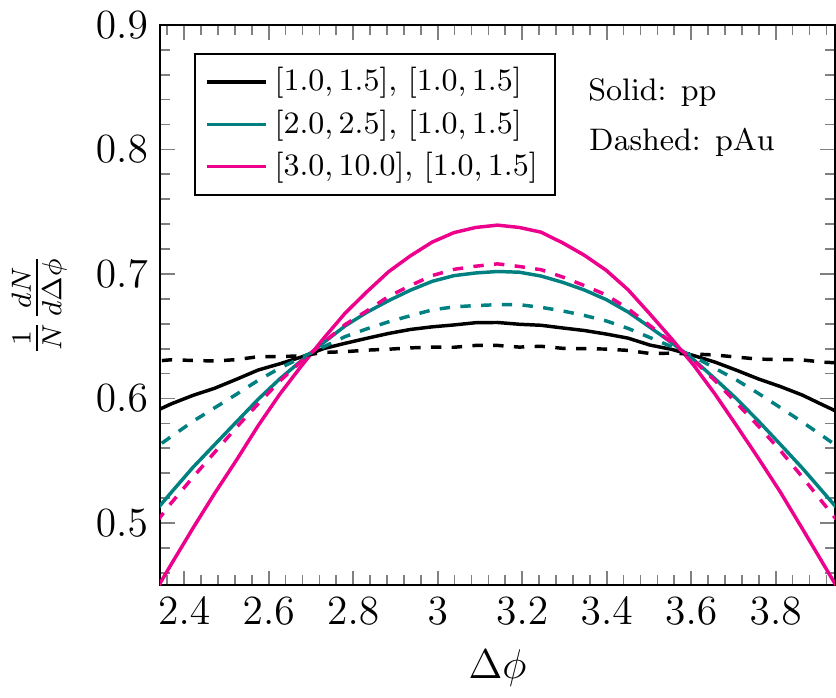}
\includegraphics[width=0.35\linewidth]{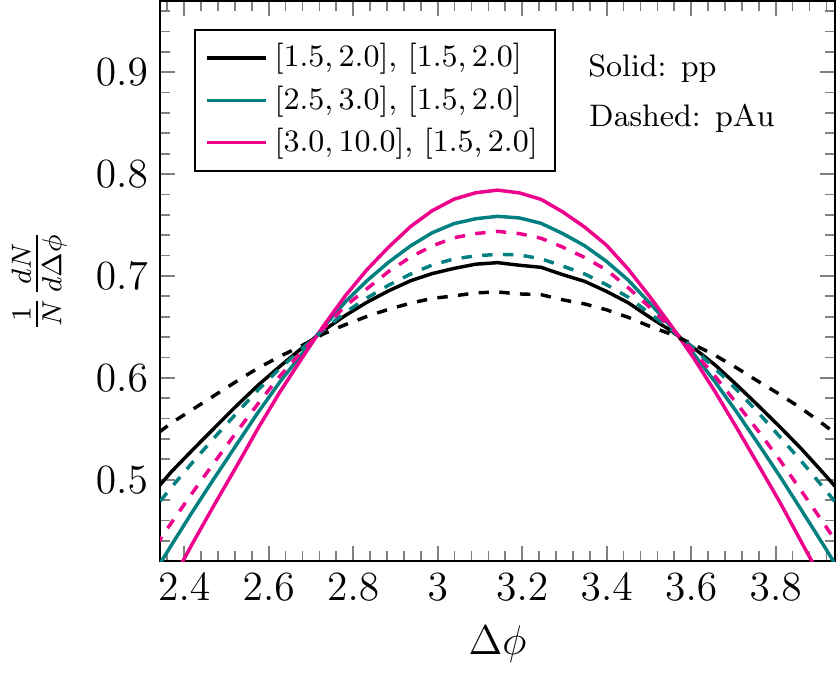}\\
\includegraphics[width=0.35\linewidth]{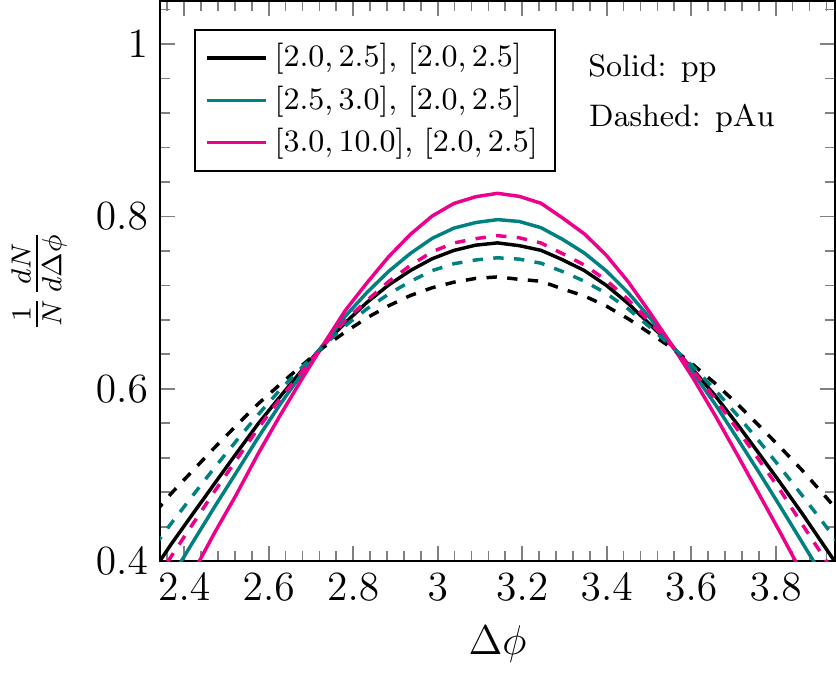}
\includegraphics[width=0.35\linewidth]{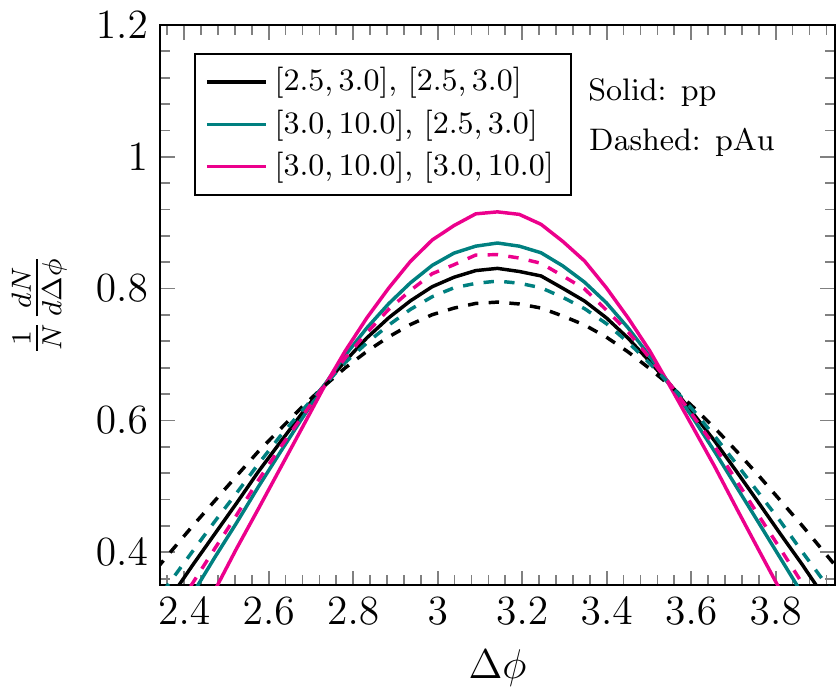}
\caption{Prediction for normalized forward dihadron ($\pi^0$) angular correlation in the forward rapidity region ($2.6<y<4.2$) in pp and minimal bias pAu collisions at $\sqrt{s}=200$ GeV. The first $p_T$ bin is for the trigger $\pi^0$, while the second bin is for the associate $\pi^0$.}
\label{pre}
\end{figure}

Finally, we make predictions for several transverse momentum bins, both for trigger and associate particles, as shown in Fig.~\ref{pre} for both $pp$ and $pAu$ collisions at RHIC. As we can see in the plots, by comparing solid (or dashed) curves with different colors, which correspond to different $p_T$ of trigger particle, we find that the correlation curves become flatter when we decrease the transverse momentum. Despite the fact that the strength of the perturbative Sudakov factor increases with $p_T$, partons with larger transverse momenta are less likely to be deflected. Therefore the resulting distribution in $p_T$ is also less likely to be  broadened. This is the reason why we see the corresponding curve of the $p_T$ bin with large transverse momentum is more steep than that of the small $p_T$ bin. Furthermore, by comparing the solid and dashed curves with the same color, we see that the back-to-back dihadrons are always more decorrelated in $pAu$ collisions than in $pp$ collisions. This is understood as originating from the  larger saturation effects in nucleus target. 

\section{Conclusions}

In this paper, we have carried out a comprehensive study of forward rapidity dihadron angular correlations in both $pp$ and $dAu$ ($pA$) collisions at RHIC, by using the small-$x$ formalism with parton shower effects. This new framework allows to describe the forward dihadron angular correlation in $pp$ collisions, where both the small-$x$ effect and the Sudakov effect are important. By incorporating the parton shower effect, a very good agreement with all the available data is obtained, and further prediction for the upcoming data collected in the $pAu$ collisions at RHIC is also provided. Using the results in $pp$ collisions as the baseline, we can reliably study the saturation effect which accounts for the difference between angular correlations the $pp$ and $pAu$ collisions, and therefore provide robust predictions. This would allow us to systematically study the signature of gluon saturation at RHIC.  


\begin{acknowledgments}
This paper is based upon work partially supported by the NSFC under Grant No.~11575070. It is also supported by the U.S. Department of Energy, Office of Science, Office of Nuclear Physics, under contracts No.  DE-AC02-05CH11231, DE-SC-0002145, by the National Science Center, Poland, Grant No. 2015/17/B/ST2/01838 and within the framework of the TMD Topical Collaboration, by the Agence Nationale de la Recherche under the project ANR-16-CE31-0019. We thank Akio Ogawa, Elke-Caroline Aschenauer and Cyrille Marquet for useful comments and correspondence. 
\end{acknowledgments}




\end{document}